\documentclass[11 pt, preprint]{revtex4}

\usepackage{epsfig}
\usepackage{amssymb}
\newcommand{\be}{\begin{equation}}
\newcommand{\ee}{\end{equation}}
\newcommand{\ba}{\begin{eqnarray}}
\newcommand{\ea}{\end{eqnarray}}
\usepackage{braket}
\begin{document}
\title{Entanglement properties of bound and resonant few-body states}



\author{Arkadiusz Kuro\'s}
\affiliation{Instytut Fizyki imienia Mariana Smoluchowskiego, Uniwersytet Jagiello\'nski, ulica Profesora S. \L ojasiewicza 11, PL-30-348 Krak\'ow, Poland
}
\affiliation{Institute of Physics,  Jan Kochanowski University\\
ul. \'Swi\c{e}tokrzyska 15, 25-406 Kielce, Poland}
\author{Anna Okopi\'nska}
\affiliation{Institute of Physics,  Jan Kochanowski University\\
ul. \'Swi\c{e}tokrzyska 15, 25-406 Kielce, Poland}

\begin{abstract}
Studying the physics of quantum correlations has gained new interest after it has become possible to measure entanglement entropies of few body systems in experiments with ultracold atomic gases. Apart from investigating trapped atom systems, research on correlation effects in other artificially fabricated few-body systems, such as quantum dots or electromagnetically trapped ions, is currently underway or in planning. Generally, the systems studied in these experiments may be considered as composed of a small number of interacting elements with controllable and highly tunable parameters, effectively described by Schr\"odinger equation. In this way, parallel theoretical and experimental studies of few-body models become possible, which may provide a deeper understanding of correlation effects and give hints for designing and controlling new experiments. Of particular interest is to explore the physics in the strongly correlated regime and in the neighborhood of critical points.

Particle correlations in nanostructures may be characterized by their entanglement spectrum, i.e. the eigenvalues of the reduced density matrix of the system partitioned into two subsystems. We will discuss how to determine the entropy of entanglement spectrum of few-body systems in bound and resonant states within the same formalism. The linear entropy will be calculated for a model of quasi-one dimensional Gaussian quantum dot in the lowest energy states. We will study how the entanglement depends on the parameters of the system, paying particular attention to the behavior on the border between the regimes of bound and resonant states.
\end{abstract}
{\bf Keywords:} correlations, entanglement entropy, quantum dots, resonances.

\maketitle

\section{Introduction}
Experimental realizations of artificial few-body systems in ultracold atomic gases or semiconductor nanostructures offer a unique opportunity to get an insight into the quantum nature of the world. The fabricated microscopic structures, such as quantum dots (QD), electromagnetically trapped few-ion systems, or clusters of ultracold atoms, are often called ,,artificial atoms''. Similarly as natural atoms, they can be theoretically well described by a nonrelativistic Schr\"odinger equation with a Hamiltonian containing the trapping potential $V(\textbf{r}_i)$ and the two-body potentials $U(|\textbf{r}_i-\textbf{r}_j|)$ of interaction between the constituents:

\begin{equation}\left[\sum ^N_{i=1}\left(-\frac{\hbar^2}{2m}\nabla^2_i+V(\textbf{r}_i)\right)+\sum ^N_{i<j} U(|\textbf{r}_i-\textbf{r}_j|)\right]\psi(\textbf{r}_1,\dots,\textbf{r}_N)=E\psi(\textbf{r}_1,\dots,\textbf{r}_N).
\end{equation}

In most cases, the ,,artificial atoms'' can be considered as closed quantum systems having a discrete spectrum of energy levels analogous to that of naturally existing few-body systems such as few-electron atoms or molecules. However, the artificially fabricated structures possess many advantages over the natural ones. The most important one is the possibility of isolating them in experiments and manipulating individually. Moreover, their microscopic parameters can be precisely tuned, not only the number of constituents, but also the type and the strength of interaction between them as well as the shape of the confining potential may be experimentally controlled by applying appropriately designed external fields.  In this way, a system described by a specific Hamiltonian may be engineered at will, which makes a comparison of its measured and calculated characteristics in dependence on varying parameters possible. In particular, the system can be driven from stable to unstable state, which allows a study of the correlations in the critical regime.

There are various ways of quantifying correlations in quantum systems. In quantum chemistry, the correlations are usually measured with respect to Hartree-Fock approximation. The Hartree-Fock picture provides a simple way of understanding the interacting system by mapping it onto a system of non-interacting particles moving in a self-consistent field of other particles. In this picture, the exchange correlations due to Fermi statistics are accounted for, since the wave function of the system is approximated by a determinant of one-electron functions, but the correlations arising from the Coulomb interaction between electrons are beyond the scope of this approach. The amount of Coulomb correlations has been discussed first by Wigner and Seitz~\cite{WignerSeitz}, who introduced a quantity called the correlation energy, defined as a difference between the exact ground-state energy and its Hartree-Fock approximation $E_{corr}=E-E_{HF}$~\cite{WignerSeitz, Lowdin}. $E_{corr}$ is frequently used until now as an energetic measure of correlation effects. Generalizing this concept, the correlation energy for states of definite permutation symmetry may be defined with respect to the mean field approximation to the energy $E_{MF}=\langle\Psi_{MF}|H|\Psi_{MF}\rangle$, where the approximate wave function is of the form of symmetrized or antisymmetrized product
\begin{equation}\Psi_{MF}(\textbf{r}_1,\dots,\textbf{r}_N)\approx \frac{1}{\sqrt{N}}\left\{
\begin{array}{ccc}
\mathcal{S}(\phi_{i_1}(\textbf{r}_1),\dots,\phi_{i_N}(\textbf{r}_N)) \mbox{~~for bosons} \\
\mathcal{A}(\phi_{i_1}(\textbf{r}_1),\dots,\phi_{i_N}(\textbf{r}_N)) \mbox{~~for fermions}\\
\end{array}
\right. .  \label{bas} \end{equation}	
However, one has to point out that so defined $E_{corr}$ is not a measurable quantity and its theoretical determination is troublesome, since it requires the mean field calculation to be performed in addition to numerical determination of the exact energy. In strict sense, $E_{corr}$ is not a measure of the true correlation strength but a measure of quality of the mean field approximation.

Another way of describing correlations in many-body systems is by using entropic concepts from quantum information theory~\cite{Amico, Tichy}. Theoretically, the bipartite correlations in pure states of many-body systems are characterised by the entanglement spectrum of the reduced density matrix (RDM) of the system partitioned into two subsystems~\cite{compendium, Nielsen}. They can be quantified by entanglement entropies, i.e. von Neumann or other Renyi entropies of that spectrum. It was conjectured by Collins~\cite{Collins} that the correlation energy is proportional to the von Neumann entropy. The original conjecture was shown to fail for the Hooke atom, since $E_{corr}$ does not vanish in the limit of vanishing interaction between constituents and its modification has been proposed~\cite{Ziesche}. The improved conjecture states that  the relative correlation energy  $\Delta E=|{E_{MF}-E_{exact}\over E_{exact}}|$ is proportional to the entanglement entropy and that is $\Delta E$ that has to be used as an energetic correlation measure~\cite{Ziesche}. The modified conjecture has been shown to hold in two-electron elliptic quantum dots~\cite{elliptic} and two-electronic He-like series~\cite{Dehesa}, by demonstrating that the relative correlation energy  $\Delta E$ is linearly related to the entanglement measures associated with the linear and von Neumann entropies of the RDM.

The advantage of entropic correlation measures is that they are defined without referring to mean field approximations and may be calculated from  numerically determined few-body wave function. Recently, the methods to measuring entanglement entropies of many body systems have been developed. The easiest to measure experimentally is the linear entropy that is related to the purity of the reduced system. The RDM purity has been measured in experiment at Harvard University by creating two copies of the four-atom system in optical lattice of controllable depth and interfering them with each other~\cite{Islam}. Measuring differences between different parts of the two systems when the whole remained identical, they were able to measure entanglement in the system. Those findings suggest that performing measurements of other kinds of entanglement entropies will be feasible by producing more copies of the many-body systems.

The aim of our research is to apply the concepts from quantum information theory to study resonance phenomena in few-body systems. In our study we develop an approach that enables determination of entanglement entropies of both bound and resonant states. As an application we study the model of the QD consisting of two interacting particles trapped in an open external potential. The linear entropy is calculated in dependence on the parameter related to the width of the potential well. The most interesting  is the threshold range, where the character of the lowest energy state changes from bound to resonant.

The plan of our work is the following. The biorthogonal formalism enabling a treatment of resonances in analogous way to bound states will be presented in Sec.\ref{Resonance}. In Sec.\ref{Spectrum} the optimized Rayleigh-Ritz method will be generalized to resonant states. In Sec.\ref{Bipartite entanglement} we discuss the quantum information characteristics of correlations in the system. Entanglement entropies in the lowest states of the Gaussian QD are calculated in Sec.\ref{Results}, and the conclusion presentd in Sec.\ref{Conclusion}

\section{Bound and resonant states}
\label{Resonance}
Recent advances in experimental fabrication of the systems that realize tunable few-particle Hamiltonians give hope that measurements of their detailed characteristics in dependence on varying parameters will be possible. Especially interesting range is that around the critical value of their parameters, which divides the region of stability from that of instability. Theoretical description of this range is not an easy task because the stable and unstable states are described quite differently in quantum mechanics. Bound state wave functions fulfil the Schr\"{o}dinger equation with vanishing boundary conditions at infinity, and belong to the Hermitian domain of the Hamiltonian. Unstable states, called resonant or quasi-bound states, have finite but relatively long lifetime, which corresponds to finite probability of decaying. Although they might exhibit localized properties, the important distinction from the bound states is that their wave functions, being solutions to the Schr\"odinger equation with outgoing boundary conditions,  diverge in the continuum. The resonant states cannot be thus described as vectors in the Hilbert space of square-integrable functions $\mathcal{L}^2$. In the following, the possibility of treating bound and resonant states on the same footing will be presented, limiting to the case of one particle in one-dimensional space for simplicity.

\subsection{Resonances as Gamow states}
Although the lifetime of resonant states is finite, they may be considered as eigenstates of the Hamiltonian operator.
In studying radioactive $\alpha$-decay, Gamow proposed~\cite{Gamow} to relate the resonant state with the solution of the Schr\"{o}dinger
equation
\begin{equation} \hat{H}\psi_{res}(x)=\varepsilon \psi_{rez}(x), \label{cse}\end{equation}
with complex eigenvalue
\begin{equation} \varepsilon= E - i\frac{\Gamma}{2}. \label{complex_ener}\end{equation}
This results in the probability density of finding the particle of the form
\begin{equation} |\psi_{res} (x,t)|^2= e^{-\Gamma t/\hbar} |\psi_{res} (x)|^2,\label{eq:zal1}\end{equation}
where the parameter $\tau=\frac{\hbar}{\Gamma}$ determines the lifetime and corresponds to Breit-Wigner distribution
\begin{equation} |\psi(\epsilon)|^2= \frac{\Gamma}{2\pi}\frac{1}{(\epsilon-E)^2+(2\Gamma)^2}. \end{equation} The complex eigenvalues $E - i\frac{\Gamma}{2}$ coincide with  the positions of the poles of S-matrix in the complex energy plane. 

The inherent difficulty of such an approach is that the wave functions are not square integrable, $\psi_{rez} \not \in \mathcal{L}^2$. The rigorous way of dealing with Gamow states needs an extension of Hilbert space to the Rigged Hilbert Space. The proper treatment of the unnormalizable states of the continuous spectrum is assured by constructing Gelfand triplet $\boldmath{\Phi}\subset\mathcal{H}\subset\boldmath{\Phi}^\times$, where $\boldmath{\Phi}$ is the space of test functions and $\boldmath{\Phi}^\times$ is the space of continuous antilinear functionals over $\boldmath{\Phi}$. The Gamow vectors are generalized eigenvectors of an extension of the considered Hamiltonian, which is self-adjoint on the Hilbert space $\mathcal{H}$, with complex eigenvalues \cite{bohmTriplet, bohmQM, gadella}. Application of the formalism is however difficult to implement in the context of realistic quantum-mechanical models. In practical calculations, other method, such as closing the system in a large enough box, or introducing complex absorbing potential are thus applied. The most widely applied method that enables treating the resonances as autonomous states of the system is the method of complex scaling which we will also use in this paper.

\subsection{Complex scaling method}
\begin{figure}[h]
\scalebox{0.35}{\includegraphics{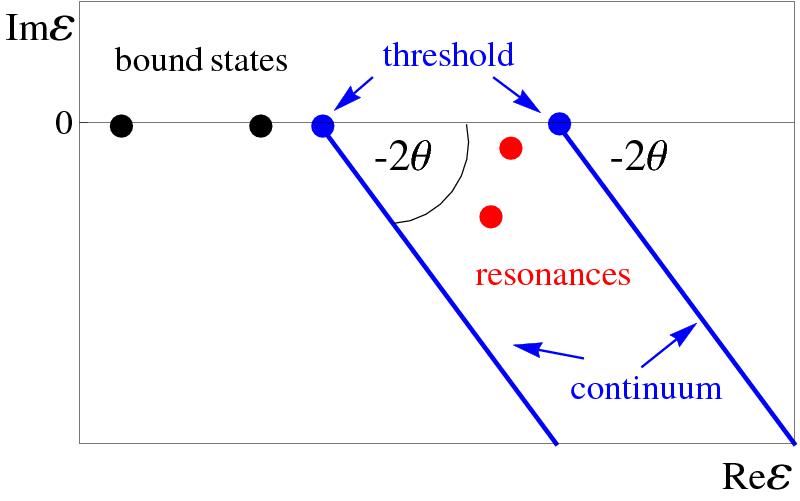}}
 \caption{The spectrum of the complex-rotated Hamiltonian $\hat{H}_{\theta}$.}
\label{widmo}
\end{figure}

The complex scaling method (CSM) allows to separate the resonant states from the continuous spectrum of the Hamiltonian $\hat{H}$ which is self-adjoint on the Hilbert space $\mathcal{L}^2$~\cite{NHQM}. Consider the scaling transformation of the form
\begin{equation} \hat{x}\to U \hat{x} U^{-1}=e^{i\theta}\hat{x}, \label{scaling}\end{equation}
where $U=e^{-\theta \hat{x} \hat{p}/\hbar}$ and $\theta \in \mathbb{R}$.
Upon the transformation the Hamiltonian takes a form
\begin{equation}\hat{H} \to U \hat{H} U^{-1}=\hat{H}_{\theta}=-e^{-2i\theta }\frac{ \hat{p}^2}{2m}+V(e^{i\theta }\hat{x}). \label{eq:hamt}\end{equation}
It is easy to observe that $\hat{H}_{\theta}$ is no more self-adjoint on $\mathcal{L}^2$, if $\theta\ne 0$. However, the advantage is that the rescaled wave functions of resonant states
\begin{equation} \psi^{\theta}_{rez}(x)=U\psi_{rez}(x)=e^{i \theta\over 2}\psi_{rez}(x e^{i \theta})
\end{equation} become square integrable if $0<\theta-\alpha_{rez}<\frac{\pi}{2}$, where  $\tan\alpha_{rez}={\Gamma\over{2 E}}$. This fact has been rigorously proved for dilatation analytic interactions, i.e. for potentials $V(x)$ analytically continuable on the complex plane, and is referred to as the Balslev-Combes theorem \cite{ABC}. After complex scaling transformation (\ref{scaling}), the energies of bound states and the thresholds remain the same as those of the original Hamiltonian $\hat{H}$, but the continuous spectra get rotated about the thresholds by an angle $2\theta$ into the lower energy half-plane, exposing complex resonance eigenvalues, as illustrated in Fig.\ref{widmo}.
As a result of this, the resonances can be determined as the eigenstates of the non-Hermitian Hamiltonian $\hat{H}_{\theta}$ by using bound-state-like strategies \cite{NHQM, ABC, Brandas, Brandas2012}. The price we pay is that we have to deal with non-Hermitian operators.

\subsection{Spectrum of a non-Hermitian operator}
The spectral analysis of non-Hermitian operators is a more complicated issue than that of the Hermitian ones. Consider a non-Hermitian operator $\hat{K}\ne \hat{K}^{\dagger}$, on a Hilbert space $\mathcal{H}$ with scalar product $\langle.|.\rangle$, where the adjoint operator is defined by $\braket{\chi|\hat{K}\psi}=\braket{\hat{K}^{\dagger}\chi|\psi}$. To simplify the treatment, we assume that the spectrum of the operator $K$ is discrete with distinct eigenvalues
\begin{equation}
\hat{K}\ket{\psi_n}=\varepsilon_n \ket{\psi_n} \ \ , \ \  \bra{\chi_n}\hat{K}= \varepsilon_n\bra{\chi_n},~n\in\mathbb{N}
\label{complexsch}
\end{equation} where the $\ket{\psi_n}$ and $\bra{\chi_n}$ eigenvectors of the operator $\hat{K}$ correspond to the same eigenvalue $\varepsilon_n$. It appears convenient~\cite{Brody} to study the intertwined eigenproblem of the adjoint operator $K^{\dagger}$ that reads
\begin{equation}
\hat{K}^{\dagger}\ket{\chi_n}=\varepsilon_n^* \ket{\chi_n} \ \ , \ \  \bra{\psi_n}\hat{K}^{\dagger}= \varepsilon_n^*\bra{\psi_n} ,~n\in\mathbb{N}.
 \label{complexschc}
\end{equation}
If the operator $\hat{K}$ is non-Hermitian, the vectors $\ket{\psi_n}$ and $\ket{\chi_n}$ are essentially different. It is customary to call $\ket{\psi_n}$ the right, and $\ket{\chi_n}$ the left eigenvector of the operator $\hat{K}$. The unpleasant feature of non-Hermitian operators is that their right eigenvectors are not orthogonal to each other ($\langle \psi_k|\psi_n\rangle \ne \delta_{kn}\langle \psi_k|\psi_n\rangle$ for $k\ne n$). Fortunately, orthogonality holds between the right and left eigenvectors that correspond to different eigenvalues, they can be thus normalized so as to satisfy the relation
\begin{equation}\braket{\chi_k|\psi_n}=\delta_{kn}, \mbox{ for all } k, n. \label{biorthonormal}
\end{equation} In this way a set $\{\ket{\chi_n},\,n\in\mathbb{N}\}$ is obtained that is biorthonormal to the set $\{\ket{\psi_n},\,n\in\mathbb{N}\}$. The choice is unique up to simultaneous multiplication of the right vectors by arbitrary complex constants $f_n$ and the left vectors by $1\over f_n^*$, which changes the norm $\sqrt{|\langle \psi_n|\psi_n\rangle|}$ by factor $|f_n|$. 
We assume the operator $\hat{K}$ is such that the completness relations hold \begin{equation}\sum_{n}\ket{\psi_n}\bra{\chi_n}=\sum_{n}\ket{\chi_n}\bra{\psi_n}=I. \label{completness}
\end{equation}
One has to mention that for non-Hermitian Hamiltonians this is not always true. It may happen for some specific values of the Hamiltonian parameters that one of its eigenvectors, e.g. the $k-$th, is such that $\braket{\chi_k|\psi_k}=0$. In such a case, which is called exceptional point \cite{Kato, Hess}, some eigenstates coalesce and completeness relations are not fulfilled. Excluding such exceptional cases, the biorthonormal set $\{\ket{\psi_n},\ket{\chi_n},n\in\mathbb{N}\}$ constitutes a Riesz basis in $\mathcal{H}$. Representation of the operator $\hat{K}$ in that basis takes the diagonal form
\begin{equation}
\hat{K}=\sum_{n}|\psi_n\rangle \varepsilon_n \langle\chi_n|.
\end{equation}
The projection operator onto the direction of $\ket{\psi_n}$ is given by
\begin{equation}
\hat{P}_{n}=|\psi_n\rangle \langle\chi_n|.
\end{equation}
The biorthonormal basis may be connected to some orthonormal basis $\{\ket{e_k}\}$ chosen in the Hilbert space $\mathcal{H}$ by the transformation
\begin{equation}A\ket{\psi_{n}}=\ket{e_{n}}, ~\ket{\chi_n}=A^\dagger \ket{e_{n}},\\\bra{e_{n}}=\bra{\chi_{n}}A^{-1}, ~\bra{\chi_n}=\bra{e_{n}}A.\end{equation} This transformation brings $\hat{K}$ to the diagonal form
\begin{equation}A\hat{K}A^{-1}=\sum_{n}\varepsilon_n A\ket{\psi_{n}}\bra{\chi_n} A^{-1}=\sum_{n}\varepsilon_n\ket{e_n} \bra{e_n}.\end{equation}
Obviously, the transformation $A$ does not have to be unitary. Only in the case of Hermitian operator, $\hat{K}=\hat{K}^{\dagger}$, the unitarity of $A$ is assured and in this case $\ket{\chi_n}=\ket{\psi_n}$ (up to a factor of unit absolute value), i.e. the biorthonormal basis reduces to the standard orthonormal basis offered by the right eigenvectors of $\hat{K}$.

Representing the vectors in biorthonormal basis $\{\ket{\psi_{n}},\ket{\chi_n}\}$ as linear combinations
\begin{equation}
\ket{\psi}=\sum_k c_{k} \ket{\psi_k}, \mbox{ and } \ket{\chi}=\sum_k d_{k} \ket{\chi_k}, \mbox{ i.e. }  \bra{\chi}=\sum_k d_{k}^*\bra{\chi_k},\label{badis}
\end{equation}
their scalar product is expressed as
\begin{equation}
\braket{\chi|\psi}=\sum_k d_{k}^* c_{k}.\label{badis}
\end{equation}
This means that the bra and ket vectors can be viewed as rows and columns, respectively.
Similarly, with the matrix element of an arbitrary operator $\hat{F}$ in the biorthonormal basis defined as
\begin{equation} F_{nk}=\bra{\chi_n}\hat{F}\ket{\psi_k},\end{equation}
the product of operators $\hat{F}$ and $\hat{G}$ is represented simply by the matrix $(\hat{F}\hat{G})_{nm}=\sum_{k}F_{nk} G_{km}$, i.e. the standard rules of matrix multiplication apply. This wouldn't be the case, if nonorthogonal basis $\{\ket{\psi_{n}}\}$ with matrix element defined as $\bra{\psi_n}\hat{G}\ket{\psi_k}$ would be used. In biorthonormal bases, the matrix algebra looks very similar to the algebra in orthonormal ones. It has been stated \cite{Brody} that for consistent probabilistic interpretation of the eigenstates of non-Hermitian Hamiltonians, the duality relation has to be modified. For an arbitrary state $\ket{\psi}$, the associated state has to be defined according to
\begin{equation}
\ket{\psi}=\sum_n c_n\ket{\psi_n} \ \ \Leftrightarrow \ \ \ket{\widetilde{\psi}}=\sum_n c_n\ket{\chi_n}.
\label{asoc}
\end{equation}
Observing that $c_n = \braket{\chi_n|\psi} = \braket{\widetilde{\psi}|\psi_n}$, it might be verified that $\braket{\widetilde{\psi}_1|\psi_2} = \braket{\psi_1|\widetilde{\psi}_2}$ and the probability for a transition from a quantum state $\psi$ to the $n$-th eigenstate of the Hamiltonian is a real number.

We consider the case when $\hat{K}$ is a Hamiltonian operator that fulfills the time dependent Schr\"odinger equation
\begin{equation}
i\hbar {\partial\over\partial t} \ket{\psi} =\hat{K}\ket{\psi}.
\label{TDS}
\end{equation}
In that case, $\hat{K}^\dagger$ and $\hat{K}$ are connected by the time reversal operator $\mathcal{T}$ by the relationship
$\hat{K}^\dagger=\mathcal{T}\hat{K}\mathcal{T}^{-1}$, and the phase can be chosen so as to have $\ket{\chi_n}=\mathcal{T}{\ket{\psi_n}}.$ For nondegenerate problems in 1D, this results in a simple relationship between the eigenvectors represented in the position space
\begin{equation}
\chi_{n}(x)=\psi_n^*(x).
\label{timereversal}
\end{equation}
With such a fixing of the biorthonormal basis, it is easy to show that $\langle \psi_n|\psi_n\rangle=r$, where $r\ge 1$, and the equality holds only in the case if $\hat{K}$ is Hermitian in $\mathcal L^2$ \cite{Rotter}.

\section{Determination of the spectrum of a non-Hermitian Hamiltonian}
\label{Spectrum}
\subsection{Rayleigh-Ritz approach}
After complex scaling, the Hamiltonian $\hat{H}_{\theta}$ becomes non-Hermitian operator in $\mathcal{L}^2$.
Determination of its spectrum may be based on the bivariational principle for non-Hermitian operators \cite{LowdinRR}. Since the resonance eigenvalues are complex numbers, their spectrum is determined by stationarity rather than minimization condition. The complex Rayleigh quotient is defined as
\be
I[\chi,\psi]=\frac{\braket{\chi|\hat{H}_{\theta}|\psi}}{\braket{\chi|\psi}}, \ \  I^*[\psi,\chi]=\frac{\braket{\psi|\hat{H}_{\theta}|\chi}}{\braket{\psi|\chi}},
\ee
where the vectors $\ket{\chi}$ and $\ket{\psi}$ are such that $\braket{\chi|\psi}\ne 0$. The principle states that $I[\chi,\psi]$ is stationary with respect to independent variation of $\ket{\chi}$ and $\ket{\psi}$ iff the vectors are solutions of the eigenequations
\begin{equation}
\hat{H}_{\theta}\ket{\psi}=\varepsilon \ket{\psi}, \ \  \hat{H}_{\theta}^{\dagger}\ket{\chi}= \varepsilon^* \ket{\chi}.
\label{complexH}
\end{equation}
In the Rayleigh-Ritz approach the unknown vectors $\ket{\chi}$ and $\ket{\psi}$ are represented in a conveniently chosen basis and the expansion coefficients are fixed so as to make the Rayleigh quotient stationary, which converts the problem to the matrix form. The simplest equations are obtained if a biorthonormal basis is used and the vectors are expanded as $\ket{\psi}=\sum_k c_{k} \ket{\psi_k}$ and $\ket{\chi}=\sum_k d_{k} \ket{\chi_k}$. The exact representation of the eigenequations (\ref{complexH}) is obtained in the form
\begin{equation}
\sum_{k}\left(\braket{\chi_n|\hat{H}_{\theta}|\psi_k}-\varepsilon \delta_{nk}\right)c_k=0,\ \  n=1,2,\ldots,
\label{secular}
\end{equation}
and
\begin{equation}
 \sum_{k}\left(\braket{\psi_n|\hat{H}_{\theta}^\dagger|\chi_k}-\varepsilon^* \delta_{nk}\right)d_k=0,\ \  n=1,2,\ldots.\end{equation}
Numerical solutions in the Rayleigh-Ritz method are derived by truncating the above infinite systems to finite sum approximations that involve only the first $M$ eigenvectors. The accuracy of the approximation may be systematically improved by increasing $M$ and checking the convergence properties.

\subsection{Complex Basis}
\label{CB}
In the position space representation, $\psi(x)$ being the eigenfunction of the complex scaled operator $\hat{H}_{\theta}$ is approximated by a finite linear combination of the real functions from the chosen orthonormal basis $\{\phi^A_m(x), m\in \mathbb{N}\}$ in $\mathcal{L}^2$, where by $A$ we denoted an arbitrary real parameter. It is an usual practice to introduce a dependence on arbitrary nonlinear parameters into the functions of the basis, which makes them adaptable to the problem under study. Of course, the exact solutions of the secular equation (\ref{secular}) do not depend on their values. Therefore, in the finite $M$ approximation the freedom in the choice of the values of those parameters may be conveniently exploited by adjusting them so as to obtain the fastest convergence.

The matrix elements in the secular equation are obtained in the form
\begin{equation}
H^{A,\theta}_{jm}=\langle \phi^A_j |\hat{H}_{\theta}| \phi^A_m \rangle=\int_{-\infty}^{\infty}\phi^A_j(x)\hat{H}_{\theta}\phi^A_m(x)dx.
\end{equation}
It has been observed~\cite{ComplexBasis,Reinhardt} that changing the variable $x$ to $x e^{-i\theta}$ and using Cauchy's theorem to distort the integration contour back to the real axis, the matrix elements turn into
\begin{equation}
H^{A,\theta}_{jm}=e^{-i\theta}\int_{-\infty}^{\infty}\phi^A_j(xe^{-i\theta})\hat{H}\phi^A_m(xe^{-i\theta})dx.\label{scale}
\end{equation}
The complex scaling is thus equivalent to working with original Hamiltonian $\hat{H}$ and using the basis functions with coordinates rescaled with $e^{-i\theta}$ factor. Going further and choosing the nonlinear parameter $A$ as the scale parameter, so that
\begin{equation}
\phi^{A}_j(x)={1\over \sqrt{A}}\phi_{j}\left({x\over A}\right),
\end{equation}
the RR matrix element (\ref{scale}) may be written as
\begin{equation}
H^{\alpha}_{jm}=\int_{-\infty}^{\infty}\phi^\alpha_j(x)\hat{H}\phi^\alpha_m(x)dx,
\label{Hcomplex}
\end{equation}
where the nonlinear parameter $A$ and the complex scaling angle $\theta$ did combine into a complex parameter $\alpha=Ae^{i\theta}$. In this way, by replacing the real parameter $A$ in the square integrable functions $\phi^A$ by a complex parameter $\alpha$, we obtain the set of complex-valued functions $\phi^\alpha$ that do not necessarily belong to $\mathcal{L}^2$.
Equation (\ref{Hcomplex}) is sometimes interpreted so that the secular equation for resonances is the same as for bound states if instead of the ordinary scalar product of the Hilbert space  $<f|g>= \int_{-\infty}^{\infty} f^{*}(x)g(x) dx$, the c-scalar product $(f|g)= \int_{-\infty}^{\infty} f(x)g(x) dx$~\cite{c-product} is used. Taking into account the relation between the functions of the biorthonormal basis (\ref{timereversal}), makes evident that this interpretation is equivalent to working with the ordinary scalar product $<f|g>$ and using biorthonormal basis, which approach we adopt in the present work.

In the complex basis approach, the value of the $\alpha$ parameter may be chosen by the trial and error or determined in iterative calculation so as to ensure a fast convergence for a particular state. Another option, $RR_{opt}$, proposed by one of us for bound states \cite{AOAO}, is to fix the values of unphysical parameters in the Mth order approximation so as to make the trace of the RR matrix stationary. As representing approximation to a physical quantity (a sum of M eigenvalues), the trace should depend as weakly as possible on infinitesimal changes of unphysical parameters. This approach has an advantage that unphysical parameters are fixed before diagonalization of the RR matrix. Extending the $RR_{opt}$ method to resonant states has been proposed by requiring stationarity of the trace with respect to small variation of the complex parameter
\begin{equation} \left(\frac{\partial }{\partial \alpha} \textrm{Tr}(\textbf{H}_{\alpha}) \right) \Big{|}_{\alpha=\alpha_{opt}}=0, \label{sta2}\end{equation}
where $\textbf{H}_{\alpha} =[H^{\alpha}_{ij}]$. Convergence of the $RR_{opt}$ method has been demonstrated for one-particle resonant problems \cite{RRoptyRes}.

\subsection{Biorthonormal basis of the HO with complex frequency}
In the case where the spectrum of the non-Hermitian operator $\hat{K}$ can be solved analytically, the system of its eigenvectors together with the eigenvectors of its adjoint $\hat{K}^\dagger$ provides an explicit realization of a biorthonormal basis. Such a basis may be applied in Rayleigh-Ritz determination of the spectra of non-Hermitian operators, in analogous way as orthonormal bases of solvable Hamiltonians eigenvectors are used to determine spectra of Hermitian operators. Biorthonormal bases may be constructed from the eigenvectors of the well-known solvable problems, such as harmonic oscillator, or particle in infinite well or Coulomb potential, by replacing the real parameters of the models by complex ones. 

In solving 1D problems, we will use the biorthonormal basis  constructed from the eigenvectors of the harmonic oscillator (HO) with complex frequency $\Omega$. In the position representation, the Hamiltonian of the HO takes a form
\be \hat{H}_{HO}=-\frac{1}{2}\frac{d^2}{dx^2}+ \frac{1}{2}\Omega^2 x^2\ee
and its eigenequation reads
\begin{equation}
  \hat{H}_{HO}\psi^{\Omega}_j(x)=\left[-\frac{1}{2}\frac{d^2}{dx^2}+ \frac{1}{2}\Omega^2 x^2\right]\psi^{\Omega}_j(x)=\varepsilon_j\psi^{\Omega}_j(x).
    \label{eq:hl1}
\end{equation}
The eigenfunctions are given by
\begin{equation} \psi^{\Omega}_j(x)=\braket{x|\psi_j^{{\Omega}}}=\left(\frac{\sqrt{\Omega}}{\sqrt{\pi}2^j j!}\right)^{1/2} H_j (\sqrt{\Omega}x) e^{-\frac{\Omega x^2}{2}},
    \label{eq:eihHO}
\end{equation} where $H_j(\sqrt{\Omega}x)$ are Hermite polynomials.
In the case of complex $\Omega$, the Hamiltonian is non-Hermitian, and the eigenequation of the adjoint operator reads
\begin{equation}
  \hat{H}_{HO}^{\dagger}\chi^{\Omega}_j(x)=\left[-\frac{1}{2}\frac{d^2}{dx^2}+ \frac{1}{2}(\Omega^*)^2 x^2\right]\chi^{\Omega}_j(x)=\varepsilon\emph{}_j^*\chi^{\Omega}_j(x).
    \label{eq:hl1}
\end{equation}
The eigenfunctions of the adjoint operator $\hat{H}_{HO}^{\dagger}$ given by
\begin{equation}
\chi^{\Omega}_j(x)=\braket{x|\chi_j^{{\Omega}}}=\psi^{\Omega^{*}}_j(x)=\left(\psi^{\Omega}_j(x)\right)^{*}, \mbox{ for all } j\in \mathbb{N}
\end{equation}
are complex conjugates of the eigenfunctions of $ \hat{H}_{HO}$ (\ref{eq:eihHO}) in agreement with the general rule (\ref{timereversal}). The functions $\psi^{\Omega}_j(x)$ and $\chi^{\Omega}_j(x)$ are orthonormal with respect to the usual scalar product $\braket{f|g}=\int f^*(x)g(x) d\,x $ in the functional space, since
\begin{equation}
\bra{\chi^{\Omega}_n}\psi^{\Omega}_k\rangle=\int \left(\psi^{\Omega^*}_n(x)\right)^*\psi^{\Omega}_k(x) d\,x=\int \psi^{\Omega}_n(x)\psi^{\Omega}_k(x) d\,x=\delta_{nk}= \left(\psi^{\Omega}_n,\psi^{\Omega}_k\right).\label{scalarprod}
\end{equation}
The matrix element of an arbitrary operator $\hat{F}$ in the biorthonormal basis can be written as
\be F_{nk}=\bra{\chi^{\Omega}_n}\hat{F}\ket{\psi^{\Omega}_k}=\int \left(\psi^{\Omega^*}_n(x)\right)^*\hat{F}\psi^{\Omega}_k(x) d\,x=\int \psi^{\Omega}_n(x)\hat{F}\psi^{\Omega}_k(x) d\,x=\left(\psi^{\Omega}_n,\hat{F}\psi^{\Omega}_k\right)\label{matrel}\ee
where the equivalent expressions in terms of $c-$product have been placed on the right sides of the formulas (\ref{scalarprod}) and (\ref{matrel}) for comparison.

\section{Bipartite entanglement}
\label{Bipartite entanglement}
\subsection{Schmidt decomposition}
Entanglement expresses the correlation between various parts of a quantum system. Convenient tools for its quantification are provided by quantum information entropies which measure the increase of entropy that occurs if a system is partitioned~\cite{Nielsen}. The study of bipartite entanglement relies on the Schmidt theorem that there exists a decomposition of the state $\ket{\psi}$ on two parts in the form
\begin{equation}
\ket{\psi}=\sum_i k_i \ket{u_i}_1 \ket{v_i}_2,
\label{schmidtwf}
\end{equation}
where $\{\ket{u_i}_1\}$ i $\{\ket{v_i}_2\}$ are vectors in Hilbert spaces $\mathcal{H}_1$ and $\mathcal{H}_2$, respectively, and $\sum |k_i|^2=1$, conforming to the normalization condition $\braket{\psi|\psi}=1$. Density operator that corresponds to the pure state $\ket{\psi}$ can be written as
\begin{equation}
\hat{\varrho}_\psi =  \ket{\psi}\bra{\psi}, 
\label{DO}
\end{equation}
and reduced density operators of the subsystems $1$ and $2$, obtained by tracing out the complementary system, are represented as
\begin{equation}
\hat{\varrho}_1 =\mbox{Tr}_2 \ket{\psi}\bra{\psi}=\sum_i \lambda_i \ket{u_i}_{1\mbox{~}1}\!{\bra{u_i}},
\label{d1}
\end{equation}
and
\begin{equation}
\hat{\varrho}_2 =\mbox{Tr}_1 \ket{\psi}\bra{\psi}=\sum_i \lambda_i \ket{v_i}_{2\mbox{~}2}\!{\bra{v_i}},
\label{d2}
\end{equation}
where $\lambda_i=|k_i|^2$ are the occupation numbers that are equal for both orbitals $\ket{u_i}$ and $\ket{v_i}$. The number of non-zero terms in the decomposition (\ref{schmidtwf}) is called the Schmidt number. The considered parts are unentangled when the Schmidt number is equal one, i.e. the system is separated into two independent subsystems. The amount of bipartite entanglement can be conveniently quantified by the R\'enyi entropies of the spectrum
\begin{equation}S^{(q)}\! =\!\frac{1}{1 - q} \ln Tr(\hat{\varrho}^q)\! =\!\frac{1}{1 - q} \ln \sum_{k} \lambda_k^q, \mbox{~where~} q\in\mathbb{N},
\label{Renyi}
\end{equation}
where the subscript of the reduced density operator $\hat{\varrho}$ is omitted, since the entropies of the subsystems $1$ and $2$ are equal.
The most used are the first and second R\'enyi entropies. The von Neumann entanglement entropy, obtained as the limit as $q\to 1$, can be represented as
\begin{equation}
S=\lim_{q\to1}S^{(q)}=-Tr[{\hat{\varrho}} \ln{\hat{\varrho}}]=-\sum_{k} \lambda_{k} \ln\lambda_{k}.
\end{equation}
The second R\'enyi entropy is related to the linear entropy $L$ by the relationship $S^{(2)}=-\ln(1-L)$. It is easy to see that
\begin{equation}L=1-Tr(\hat{\varrho}^2)=1-\sum_{k}\lambda_{k}^{2},\end{equation} where $Tr[\hat{\varrho}^2]$ is the purity of the RDM.

Various kinds of entanglement can be discussed by studying different partitions of the system. Here we consider entanglement between the particle partitions, where the system is divided on two parts: $p$-particle system and $(N-p)$-particle system.
The $p$-particle RDM defined as~\cite{RDM}
\begin{equation}\!\!\varrho_{(p)}(\textbf{r}_{1}\!,\!...,\!\textbf{r}_{p},\!\textbf{r'}\!_{1},\!...,\!\textbf{r'}\!_{p})\!=\!\!\int\!\!\Psi(\textbf{r}_{1},\!...,\!\textbf{r}_{p},\!\textbf{r}_{p+\!1}\!,\!...,\!
\textbf{r}_{N})\Psi^{*}(\textbf{r'}\!_{1}\!,\!..\!,\textbf{r'}\!_{p},\!\textbf{r}\!_{p+\!1},\!...,\!\textbf{r}\!_{N})d^{3}\!r_{p+\!1}...d^{3}\!r_{N},
\end{equation}
can be represented in the Schmidt form
\begin{equation}\varrho_{(p)}(\textbf{r}_{1},\!...,\textbf{r}_{p},\textbf{r'}_{1},\!...,\textbf{r'}_{p})\!=\sum_{k}\lambda_{k}^{(p)} u_{k}(\textbf{r}_{1},...,\textbf{r}_{p}) u_{k}^{*}(\textbf{r'}_{1},...,\textbf{r'}_{p}).\label{schmidt}
\end{equation}
It is interesting to note that the linear entropy of the particle bipartition can be calculated directly from wave functions. Other Renyi entropies require determination of natural occupancies, so that diagonalisation of the RDM has to be performed. It turns out that the linear entropy is the easiest entropy to determine both theoretically and experimentally.

\subsection{Two-particle case}
In this work, we limit our test examples to systems of two particles in one-dimensional potential. In this case, the only possible partition is into two one-particle systems with RDM of simple one-particle form
\begin{equation}\!\!\varrho(x_{1}\!,\!x'_{1})\!=\!\int\!\!\Psi(x_{1},x_{2})\Psi(x'_{1},x_{2})dx_{2},
\end{equation}
the eigenfunctions of which are just the natural orbitals, well known in quantum chemistry~\cite{LowdinNO}  
and in the Schmidt form is represented as
\begin{equation}\varrho(x_{1}\!,\!x'_{1})\!=\sum_{k}\lambda_{k} u_{k}(x_{1}) u_{k}(x'_{1}),
\label{schmidt2}
\end{equation}
where we omitted the index $p=1$ of $\varrho$ and we assumed that wave functions are chosen to be real.

The entanglement in two-particle systems in various external potentials has been intensively studied by calculating linear and von Neumann entropies of RDM both for natural atoms \cite{Kais, Manzano, DehesaHe, Benetti, Ho, my} and artificial systems \cite{Ziesche, Hooke, Amovilli, Nagy2004, elipt, Yanez, MoshPKAO, Nagy2013, B-O}. In the case of resonant states, the complex scaled Hamiltonian is non-Hermitian and its eigenequation
\begin{equation}
\hat{H}_{\theta}\ket{\Psi}= \varepsilon \ket{\Psi},
\end{equation}
is intertwined with that of the adjoint operator $H^{\dagger}$ that reads
\begin{equation}
\hat{H}_{\theta}^{\dagger}\ket{X}= \varepsilon^* \ket{X}.
\end{equation}
There is some arbitrariness concerning the definition of the density operator due to the difference between the right $\ket{\Psi}$ and  left $\ket{X}$  eigenvectors of the non-Hermitian Hamiltonian. We consider two possible definitions of the density operator
\begin{equation}
\rho_\Psi={\ket{\Psi}\bra{X}\over\braket{X|\Psi}}
\label{ro}
\end{equation} 
or
\begin{equation}
\widetilde{\rho}_\Psi={\ket{\Psi}\bra{\widetilde{\Psi}}\over\braket{\widetilde{\Psi}|\Psi}},
\label{rot}
\end{equation}
where $\ket{X}$ is the eigenvector of the adjoint operator $H^{\dagger}$ and $\ket{\widetilde{\Psi}}$ is the vector associated to $\ket{\Psi}$ defined by (\ref{asoc}). Both definitions tend to the usual definition in the Hermitian limit, when the vectors $\ket{\Psi}$ and $\ket{X}$ become equal. The Schmidt decompositions of the two-particle states are related as
\begin{equation}
\Psi(x_1,x_2)=\sum_i k_i u_i(x_1) v_i(x_2),
\end{equation}
\begin{equation}
X(x_1,x_2)=\sum_i k_i^* \chi_i(x_1) \eta_i(x_2),
\end{equation}
\begin{equation}
\widetilde{\Psi}(x_1,x_2)=\sum_i k_i \chi_i(x_1) \eta_i(x_2).
\end{equation}
Taking into account that $\chi_i(x)=u_i^*(x)$ and $\eta_i(x)=v_i^*(x)$, we can see that the RDM obtained from (\ref{ro}) can be written as 
\begin{eqnarray}
\rho(x_{1},x_{2})\! &=& \mathcal{N}^{-1}\!\!\int\!\!\Psi(x_{1},x_{3}) X^{*}(x_{2},x_{3})dx_{3}\!=\mathcal{N}^{-1}\!\!\int\!\!\Psi(x_{1},x_{3}) \Psi(x_{2},x_{3})dx_{3}= \nonumber \\
&=&\mathcal{N}^{-1} \sum_i k_i^2 u_i(x_1) u_i(x_2)=\mathcal{N}^{-1}\sum_i k_i^2 v_i(x_1) v_i(x_2),
\label{rofrompsi}
\end{eqnarray}
where $\mathcal{N}=\braket{X|\Psi}=\sum_i k_i^2$. The RDM obtained from  the definition (\ref{ro}) is the same as proposed by Pont and coworkers~\cite{Pont}, who used the c-product in the Hilbert space of complex scaled functions. With such a definition, the coefficients in the Schmidt decomposition of RDM $\lambda_{i}={k_i^2\over \sum_j k_j^2}$ are complex numbers, which results in complex-valued entanglement entropies. So defined linear entropy 
\begin{equation}
L=1-Tr(\hat{\varrho}^2)=1-{\sum_{i} k_{i}^{4}\over(\sum_i k_i^2)^2} ,
\label{Lreal}
\end{equation}
was discussed for spherically symmetric two-electron QD~\cite{Pont}, adopting the interpretation of its real part as the physical quantity and the imaginary part as the uncertainty of its measurement~\cite{NHQM}. We also used that definition calculating the linear entropy of the quasi-one dimensional Gaussian QD~\cite{soft}. Derivation of the Schmidt decomposition of RDM from the Schmidt decomposition of resonant wave functions $\Psi(x_1,x_2)$ (\ref{rofrompsi})has been performed before by orthogonalization of the right vectors basis in the complex scaling formalism ~\cite{koscik} and used to determine the complex entropies in autoionizing states of the He atom~\cite{HeExc}.\\
On the other hand, the RDM obtained from (\ref{rot}) can be written as
\begin{eqnarray}
\widetilde{\rho}(x_{1},x_{2})\! &=& \widetilde{\mathcal{N}}^{-1}\!\!\int\!\!\Psi(x_{1},x_{3}) \widetilde{\Psi}^{*}(x_{2},x_{3})dx_{3}\!= \widetilde{\mathcal{N}}^{-1}\sum k_i k_i^* u_i(x_1) u_i(x_2)= \\
&=& \widetilde{\mathcal{N}}^{-1}\sum k_i k_i^* v_i(x_1) v_i(x_2)  \nonumber,
\end{eqnarray}
where $\widetilde{\mathcal{N}}=\braket{\widetilde{\Psi}|\Psi}=\sum_i |k_i|^2$. With this definition of the density matrix, the occupancies \\ $\lambda_{i}={|k_i|^2\over \sum_j |k_j|^2}$ are real numbers, which results in real-valued entanglement entropies. The linear entropy is given by
\begin{equation}
\widetilde{L}=1-{\sum_{i} |k_{i}|^{4}\over\left(\sum_i |k_i|^2\right)^2}.
\label{Lcomplex}
\end{equation}

\section{Results for quasi-one-dimensional Gaussian QD}
\label{Results}
\begin{figure}[h]
\scalebox{0.25}{\includegraphics{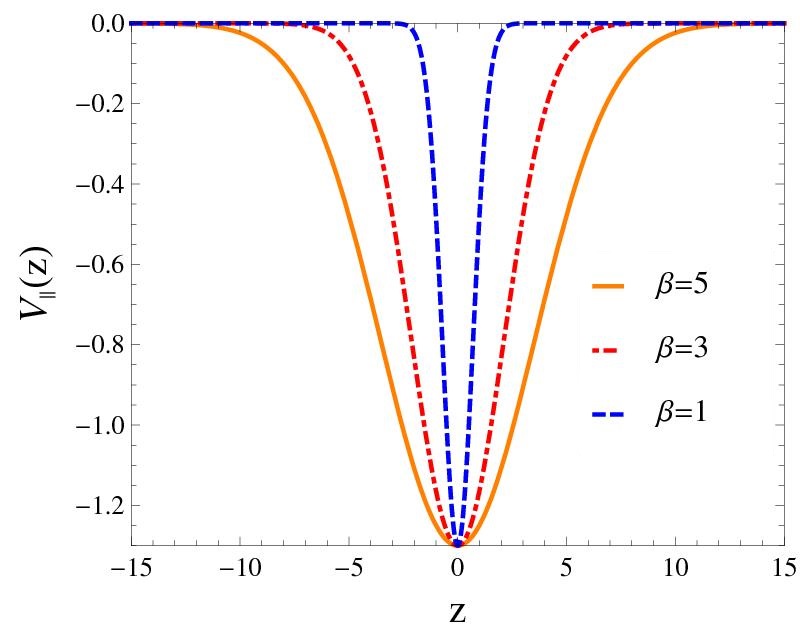}}
 \caption{The longitudinal potential for three different values of $\beta$.}
\label{pot_gauss}
\end{figure}

As an illustration, we present the entanglement entropies for the one-dimensional Hamiltonian 
\begin{equation}
\hat{H}_{1D}=\sum_{i=1}^{2}\left[-\frac{1}{2}\frac{\partial^2}{\partial
z_i^2}-V_{0}e^{-\frac{ z_i^2}{\beta^2}}\right] +V_{eff}(|z_1-z_2|).
 \label{hameff}
\end{equation}
The model can be regarded as a quasi-one dimensional approximation of the highly elongated QD consisting of two Coulombically interacting electrons, the Hamiltonian of which is given by
\begin{equation}
\hat{H}=\sum_{i=1}^{2}\left(-\frac{\hbar^2 \nabla_i^2}{2m^*} + \frac{m^* \omega_{\perp}^2}{2}(x^2_i+y^2_i)+ \\ V_{\parallel}(z_i)\right) + \frac{e^2}{4\pi \epsilon^* |\textbf{r}_1-\textbf{r}_2|} 
\end{equation}
with $m^*$ and $\epsilon^*$ being the effective electron mass and dielectric constant, respectively, which characterize the superconducting material~\cite{QD,gaus-eff}. The lateral confinement in axially symmetric harmonic potential of frequency $\omega_{\perp}$ corresponds to the lateral confinement length $\ell_{\perp}=(\frac{\hbar}{m^* \omega_{\perp}})^{\frac{1}{2}}$. With the lengths scaled to the unit of the effective Bohr radius  $a^*=\frac{4 \pi \epsilon^*\hbar^2}{m^* e^2}$, and the energies to the unit of the effective hartree energy $\mbox{Ha}^*=\frac{\hbar^2}{m^* a^2}$,  the Hamiltonian reads as
\begin{equation}
\hat{H}=\sum_{i=1}^{2}\left(-\frac{ 1}{2} \nabla_i^2 + \frac{1}{2 l_\perp^4}(x^2_i+y^2_i) - V_0 e^{-\frac{z_i^2 }{\beta^2}} \right) + \frac{1}{ |\textbf{r}_1-\textbf{r}_2|}, 
 \label{3Dham}
\end{equation}
where an attractive Gaussian potential of the depth $V_0$ expresses the longitudinal confinement with the parameter $\beta$ related to the longitudinal radius of the QD, as demonstrated in Fig.\ref{pot_gauss}. In the case of strong lateral confinement, $\ell_{\perp}~\ll~ \left(\frac{\beta^2}{2 V_0}\right)^\frac{1}{4}$, the Coulomb interaction is a small perturbation for transverse degrees of freedom. It may be thus approximately assumed that the particles stay in the lowest energy state of the transverse Hamiltonian and the excitations occur only in the longitudinal direction. The system can be approximately described by one-dimensional Hamiltonian (\ref{hameff}) with the effective electron-electron interaction in the longitudinal subspace obtained through averaging the 3D Coulomb potential over the transverse degrees of freedom in the form \cite{gaus-eff}
\begin{equation}
V_{eff}(|z_1-z_2|)=\sqrt{\frac{\pi}{2\ell_{\perp}^2}}erfcx \left[\frac{ |z_2-z_1| }{l_\perp \sqrt{2}}\right],
 \label{erf}
\end{equation}
where $erfcx (z)=\exp{x^2}(1-erf(z))$ and the error function $erf(z)={2\over \sqrt{\pi}}\int_0^x dt e^{-t^2}$. The nice feature of the effective interaction potential is its dilatation analyticity. We have checked that single mode approximation works well for the Gaussian QD at sufficiently small lateral confinement length, $\ell_{\perp}$. In this approximation the reduced density operator factorises to the form $\varrho=\varrho_{\parallel}\varrho_{\perp}$, where $\mbox{Tr}\varrho_{\perp}^2=1$, as the transverse degrees of freedom are separable. It is thus sufficient to determine the entanglement entropy from the longitudinal RDM.  

The correlations between the electrons in the highly elongated QD will be discussed thus in the quasi-one dimensional approximation (\ref{hameff}) with effective interaction (\ref{erf}). The inverse Gaussian potential has the form of an open well, the width of which depends on the value of the parameter $\beta$. The energetically lowest two-particle state is bound if its energy is smaller than the energy of the lowest one-particle state, which takes place when the width of the well is large enough. The analysis shows that there exists a threshold value $\beta_{th}$ such that for $\beta < \beta_{th}$ the lowest energy state becomes autoionizing resonant state. In order to solve the Schr\"{o}dinger equation using the same approach in the whole parameter range, we applied the complex basis method described in Sec.\ref{CB} with single particle-basis eigenfunctions of the harmonic oscillator of frequency $\Omega$ (\ref{eq:eihHO}). Allowing the nonlinear variational parameter to be complex number, enabled determination of both the bound and resonant states by diagonalization of the truncated Hamiltonian matrix $[\hat{H}_{\Omega}]_{M \times M}$ with the value of $\Omega_{opt}$ fixed by the optimization condition (\ref{sta2}). The Schmidt expansion of so determined lowest state wave function has been performed and the occupations of the RDM derived from it. With the density operator defined as (\ref{ro}), complex occupations $\lambda_{i}={k_i^2\over \sum_j k_j^2}$ and a complex linear entropy $L(\rho)$ (\ref{Lcomplex}) have been obtained, whereas the definition (\ref{rot}) resulted in real occupancies $\lambda_{i}={|k_i|^2\over \sum_j |k_j|^2}$, and a real linear entropy $L(\widetilde{\rho})$ (\ref{Lreal}). We have checked that in vicinity of $\Omega_{opt}$ the dependence of entropies on the value of $\Omega$ is slight, which justifies the results. 

\begin{figure}[h]
\scalebox{0.4}{\includegraphics{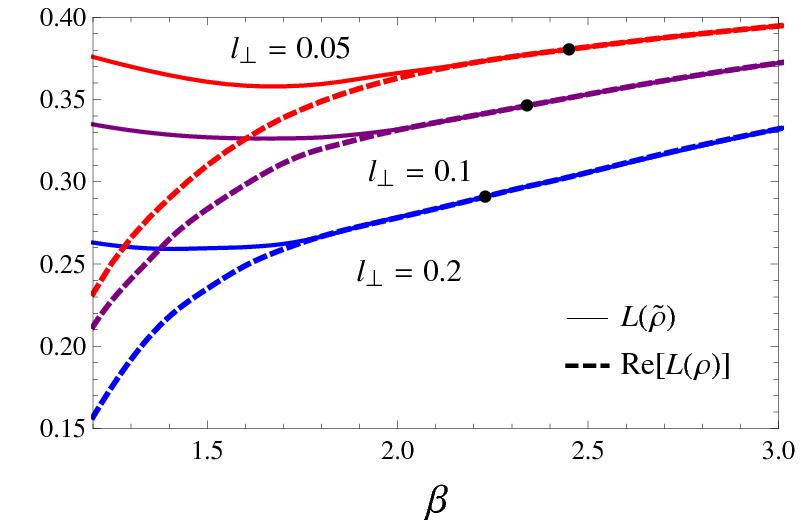}}
 \caption{Linear entropy of the quasi-1D Gaussian QD.}
\label{one}
\end{figure}

In Fig.\ref{one} the real linear entropy $L(\widetilde{\rho})$  is compared with the real part of $L(\rho)$ as function of the width of the longitudinal trap $\beta$ for several values of the lateral confinement length $\ell_{\perp}$. The critical values of $\beta_{th}$ that correspond to the ionization thresholds are marked by dots. For increasing $\beta$ the linear entropies increase, which means that electrons are more correlated in wide traps, where the trapping potential is weak in comparison with the long-range Coulomb interaction. We may observe that the entropies decrease with increasing $\ell_{\perp}$, i.e. when the transverse confinement gets weaker.  Note that the narrower the trap, the stronger is the influence of the lateral confinement $\ell_{\perp}$ on entanglement. Above thresholds the entropies are equal, and they do not differ much for the values of $\beta$ slightly below the threshold, both functions being continuous at $\beta=\beta_{th}$. The qualitative difference appears when the longitudinal trap gets narrower. The real entropy $L(\widetilde{\rho})$ gets minimum and starts to increase with decreasing $\beta$. This may be explained by the resonant character of the state. The system gets more correlated, since the probability that one of the electrons is  outside the trap gets larger and long-range Coulomb interaction dominates. It seems that real entropy better accounts for the fact that in this range of $\beta$ the number of occupied natural orbitals grows. 

\section{Conclusion}
\label{Conclusion}
We have discussed particle correlations in few-body systems subject to an open external potential. The optimized configuration interaction  method was applied to solve the Schr\"odinger equation in bound and resonant regime within the same formalism. Two possible definition of the density operator have been considered, and the RDM have been discussed for both cases. One definition resulted in complex occupancies of natural orbitals and complex linear entropy $L({\rho})$ and the other provided real linear entropy $L(\widetilde{\rho})$. The results were calculated for the model of quasi-one-dimensional Gaussian quantum dot. For the lowest energy states, the real linear entropy $L(\widetilde{\rho})$ was compared with the real part of $L({\rho})$ as functions of the parameter $\beta$ that is related to the width of the external potential well. Both functions appear continuous at the critical value $\beta=\beta_{cr}$, which separates the range where the lowest state of the system is bound from that where this state is autoionizing. However,  their behavior in the resonant regime is very different. The real part of the linear entropy monotonously decreases with diminishing with of the trap, but the real linear entropy increases indicating the growing correlation. The subject requires a broader analysis including other Hamiltonian models and will be treated in more detail elsewhere.

\section{Acknowledgements}
Support of the National Science Centre, Poland under QuantERA programme No. 2017/25/Z/ST2/03027 (A.K.) is acknowledged.

\end{document}